# Sailing towards the stars close to the speed of light

André Fűzfa,[*] Williams Dhelonga-Biarufu, and Olivier Welcomme
*Namur Institute for Complex Systems (naXys), University of Namur, Rue de Bruxelles 61, B-5000 Namur, Belgium*

We present a relativistic model for light sails of arbitrary reflectivity and absorptance undertaking nonrectilinear motion. Analytical solutions for a constant driving power and a reduced model for straight motion with an arbitrary sail's illumination are given, including for the case of a perfectly reflecting light sail examined in earlier works. When the sail is partially absorbing incoming radiation, its rest mass and temperature increase, an effect completely discarded in previous works. It is shown how sailing at relativistic velocities is intricate due to the existence of an unstable fixed point, when the sail is parallel to the incoming radiation beam, surrounded by two attractors corresponding to two different regimes of radial escape. We apply this model to the Starshot project by showing several important points for mission design. First, any misalignment between the driving light beam and the direction of the sail's motion is naturally swept away during acceleration toward relativistic speed, yet leads to a deviation of about 80 A.U. in the case of an initial misalignment of 1 arc sec for a sail accelerated up to $0.2c$ toward Alpha Centauri. Then, the huge proper acceleration felt by the probes (of order 2500 g), the tremendous energy cost (of about 13 kt per probe) for poor efficiency (of about 3%), the trip duration (between 22 and 33 years), the temperature at thermodynamic equilibrium (about 1500 K), and the time dilation aboard (about 160-days difference) are all presented and their variation with the sail's reflectivity is discussed. We also present an application to single trips within the Solar System using double-stage light sails. A spaceship of mass 24 tons can start from Earth and stop at Mars in about seven months with a peak velocity of 30 km/s but at the price of a huge energy cost of about $5.3 \times 10^4$ GW h due to extremely low efficiency of the directed energy system, around $10^{-4}$ in this low-velocity case.

## I. INTRODUCTION

The discovery of the cruel vastness of space beyond the Solar System is rather recent in human history, dating back from the measurement of the parallax of the binary star 61 Cygni by Bessel in 1838 [1]. This was even pushed orders of magnitude further by the work of Hubble on intergalactic distances in the 1920s [2,3]. Although not theoretically impossible, interstellar travel has ever been largely considered unfeasible, for a variety of good reasons. The major drawback lies in the huge gap between interplanetary and interstellar distances, initially settled by Bessel's discovery: the distance to the nearest star system, Alpha Centauri, is roughly 10 000 times the distance to planet Neptune, on the outskirts of the Solar System. It took about 40 years for the fastest object ever launched by humans, the Voyager space probe, to reach the edges of the Solar System $18 \times 10^9$ km away, at a record cruise velocity of 17 km/s. Millenia-long trips would then be needed to cross interstellar distances. Since Bessel's epoch, we know that the gap to the stars lies in these four orders of magnitude, needless to mention the much larger gap to the galaxies.

There have been many suggestions to go across the stars, and we refer the reader to [4] for an overview, some of which could even be considered as plausible, yet unaffordable, while some others are simply nonphysical. One could distinguish four categories among the (many) proposals for achieving interstellar travel: relativistic reaction propulsion, generation ships, spacetime distortions, and faster-than-light travels. We claim that only the first category is relevant for plausible discussions, but giving our arguments here would go beyond the scope of this paper. Our present paper focuses on what is perhaps the most plausible proposal among the first category: directed energy propulsion. This last is based on radiation pressure and consists of using the impulse provided by some external radiation or particle beam to propel the space ships. According to many authors [4–14], this is maybe the most promising one for three reasons: (1) it does not require embarking any propellant; (2) it allows reaching higher velocities than rockets expelling mass and submitted to the constraints of the Tsiolkovsky equation; and (3) it benefits from a strong theoretical and technical background including successful prototypes [14]. However, maybe the major drawback of directed energy propulsion lies in its weak efficiency: the thrust imparted by a radiation beam illuminating some object with power $P$ is of order $P/c$. Roughly speaking, 1 N of thrust requires an illumination on the sail of at least 300 MW.

The idea of using the radiation pressure of sunlight to propel reflecting sails dates back to the early ages of astronautics and was suggested by Tsander in 1924 (see [5] for a review of the idea). Several proposals have been put forth for using

---

solar sails for the exploration of the inner Solar System (for the first proposals, see [15,16]), to reach hyperbolic orbits with the additional thrust provided by a sail's desorption [17], or even to test fundamental physics [18]. Space probes like Ikaros and NanoSail-D2 have successfully used radiation pressure from sunlight for their propulsion. However, since solar illumination decreases as the square of the distance, this method is interesting for exploring the inner Solar System but not for deep space missions. The first realization of a laser in 1960 really opened the way to consider using them in directed energy propulsion, an idea first proposed by Forward in 1962 [6]. Since laser sources are coherent sources of light with large fluxes, one can consider sending energy over large distances toward a space ship. The Hungarian physicist Marx proposed in a 1966 paper [19] the same idea independently of Forward, together with the first relativistic model describing the straight motion of such a laser-pushed light sail. This paper was quickly followed by another one by Redding [20] in 1967 that corrected one important mistake made by Marx in the forces acting on the light sail. Twenty-five years later, Simmons and McInnes revisited Marx's one-dimensional model in [21], extending its model to variable illuminating power, examining the efficiency of the system and and how recycling the laser beam with mirrors could increase it.

Forward's idea of laser-pushed light sails for an interstellar journey has seen a strong renewal of interest since 2009 through successive funded research programs that are still active. As early as 2009, the joint NASA–University of California Santa Barbara Starlight [22] investigated the large scale use of directed energy to propel spacecrafts to relativistic speeds, including small wafer scale ones. In 2016, the Breakthrough Starshot Initiative [23] was initiated to focus on wafer scale spacecrafts and interstellar fly-bys to Alpha Centauri with the objective of achieving it before the end of the century. A nice review on large scale directed energy application to deep space exploration in the Solar System and beyond, including many detailed engineering aspects, can be found in [14].

In the context of these research programs, several authors have started modeling Solar System and interstellar missions based on this concept. Some basic one-dimensional modeling of directed energy propulsion, accounting for some relativistic effects as well as prospective applications for Solar System exploration, can be found in [24]. In [7], one can find a model for a single trip from Earth to Mars with a microwave beam-powered light sail but they did not give there the details of their model. The authors of [9–11,13] based their results on the model by Marx [19], Redding [20], and Simmons and McInnes [21], which is only valid for one-dimensional (rectilinear) motion of perfectly reflecting light sails. The case of arbitrary reflectance has not been considered so far: this involves inelastic collisions between the propulsive radiation beam and the light sail, leading to an increase of the rest mass of the last, as was already shown in [12,25] for different types of photon rockets. The authors of [9,11] provided interesting modeling of the early acceleration phases, nonrelativistic regime, and power recycling in the rectilinear motion case but did not provide the analytical solutions in terms of the rapidity as we do here. Reference [13] started from the Marx–Redding–Simmons-McInnes model and investigated the possibility to use high-energy astrophysical sources to drive the light sails during their interstellar journey. Finally, [10] provided a more complete and realistic physical model of a light sail's rectilinear motion, including notably thermal reemission of absorbed radiation through Poynting-Robertson effect and a model for development costs. Unfortunately, this author also only considered rectilinear motion and assumed the sail's thermodynamic equilibrium (constant rest mass) all along the acceleration. We will complete this model here by giving the general equations of nonrectilinear motion accounting for nonperfect reflectivity, partial absorption of incoming radiation, Poynting-Robertson drag, the sail's internal energy and temperature increase, and relaxation towards thermodynamic equilibrium. In addition, since nonrectilinear motion has not been considered so far, the effect of a misalignment of the incoming beam and the sail's velocity on the trajectory has not been investigated yet. This paper will propose a relativistic model for nonrectilinear motion of light sails with arbitrary power, reflectivity, and absorptance, investigating their general dynamics and providing useful applications to the acceleration phase of fly-by missions to outer space or single trips within the Solar System using a double-stage light sail.

In addition, the motion of mirrors under radiation pressure, possibly up to relativistic velocities, is a wide and important topic in physics. This topic can be dated back to Einstein's famous paper [26] that established special relativity and, particularly relevant here, relativistic Doppler effect, aberration, and radiation pressure. The nonrelativistic motion of moving mirrors under radiation pressure is central for optical interferometry [27,28] and its high-precision applications among which is the outstanding detection of gravitational waves. In plasma physics, thin shells of electrons and ions can constitute relativistic mirrors that can be accelerated up to relativistic velocities by the intense electromagnetic waves of a laser pulse, with the possible application to the realization of compact sources of hard electromagnetic radiation (see, for instance, [29,30]). However, these "mirrors" are nonrigid, unlike the solid sails described here. In [29], the relativistic straight motion of these deformable emergent structures of charged particles is studied both through statistical mechanics and through the particle's relativistic motion in the field of plane electromagnetic waves [31]. Our general formalism for the nonrectilinear motion of rigid mirrors under radiation pressure introduced in this paper might therefore be also of interest for applications to these relativistic mirrors in plasma physics.

In Sec. II, we recall some fundamental properties of photon rockets in special relativity, rederiving the relativistic rocket equation, and apply them to light sails. We then focus on the general motion of perfectly absorbing "white" sails and give analytical solutions for the particular case of straight motion. Then, the case of a perfectly absorbing "black" light sail is discussed and a model accounting for radiation pressure of the incoming beam, sail heating, and Poynting-Robertson drag is established. Finally, we show how one can combine the previous cases of white and black sails to build a model for the general motion of "gray" light sails with arbitrary reflectivity, and useful analytical solutions for straight motion are provided. In Sec. III, we apply previous results to (i) a dynamical system analysis of sailing at relativistic velocities, showing how intricate this discipline will be; (ii) the acceleration phase of the Starshot mission, providing many related physical quantities such as the sail's inclination, distance, proper velocity and acceleration, internal energy variation, temperature, time dilation aboard, efficiency, and trip duration; and (iii) single trips in the Solar System with Forward's idea of a multistage light sail [5]. We finally conclude in Sec. IV with some emphasis on the importance of the presented results and by giving some perspectives on this paper.

## II. GENERAL MOTION OF RELATIVISTIC LIGHT SAILS

A light sail is a spacecraft propelled by the radiation pressure exerted on its reflecting surface by some incident light beam. In the case of directed energy propulsion, this radiation is provided by some external power source, like an intense terrestrial laser. We are interested in determining the trajectory of the sail as seen from the reference frame of the external power source. Actually, this a geometrical problem of special relativity: find the sail's world line $\mathcal{L} \equiv [X^\mu(\tau)]_{\mu=0,...,3} = (cT(\tau), X(\tau), Y(\tau), Z(\tau))$ [with $(cT, X, Y, Z)$ Cartesian coordinates in the laser's frame and $\tau$ the sail's proper time] that satisfies the following equations of motion:

$$c\frac{dp^\mu}{dT} = F^\mu \quad (1)$$

where $p^\mu$ and $F^\mu$ are, respectively, the sail's four-momentum and the four-force (in units of power) acting on it in the laser's frame and $T$ is the source's proper time. It will later be useful to consider the equations of motion expressed in terms of the sail's proper time:

$$c\dot{p}^\mu = f^\mu \quad (2)$$

where a dot denotes a derivative with respect to proper time $\tau$ and the four-force in these units is given by $f^\mu = \gamma F^\mu$ where

$$\gamma = \frac{dT}{d\tau}$$

is the well-known Lorentz factor accounting notably for time dilation between the sail and the source. To determine the world line $\mathcal{L}$, one needs to remember that the sail's four-momentum $p^\mu$ is related to the tangent vector $\lambda^\mu \equiv \frac{dX^\mu}{cd\tau}$ and the sail's rest mass by $p^\mu = m(\tau)c\lambda^\mu(\tau)$, in general with variable rest mass $m$. The tangent vector is a unit timelike four-vector in spacetime, $\lambda_\mu \lambda^\mu = -1$ [with a signature $(-, +, +, +)$ for Minkowski's metric], the derivative of which, the four-acceleration $\dot{\lambda}^\mu$, is orthogonal to it: $\lambda_\mu \dot{\lambda}^\mu = 0$; in other words, the four-acceleration $\dot{\lambda}^\mu$ is a spacelike vector. The sail's rest mass $m(\tau)$ is defined by the norm of the four-momentum, $p_\mu p^\mu \equiv -m^2(\tau)c^4$, and is not constant under the action of a nonpure force ($\lambda_\mu f^\mu \neq 0$). To emphasize this important feature, we can make use of the relations above to reformulate the equations of motion (2) as the following system:

$$\begin{cases} \dot{m}c^2 = -\lambda_\mu f^\mu \\ mc^2 \dot{\lambda}^\mu = f^\mu + (\lambda_\alpha f^\alpha)\lambda^\mu \\ \dot{X}^\mu = c\lambda^\mu \end{cases} \quad (3)$$

which are nonlinear in $\lambda^\mu$. Therefore, it is obvious that a constant rest mass requires the four-force $f^\mu$ to be everywhere orthogonal to the tangent vector $\lambda^\mu$ (pure force) which includes the free motion ($f^\mu = 0$) as a trivial particular case. In general, the rest mass is therefore not constant as is the case when absorption of radiation by the sail occurs. We also refer the reader to [12,25] for several models of photon rockets with varying rest mass.

Another important additional property arises when one considers photon rockets [12], i.e., when the driving power $f^\mu$ is provided by some incoming or outgoing radiation beam, which corresponds to absorption and emission photon rockets, respectively. A light sail is actually a combination of both cases. During absorption and emission processes, the total four-momentum $p_{\text{tot}}^\mu = p^\mu + P^\mu$ of the system rocket $(p_\mu)$ and beam $(P^\mu)$ is conserved: $\dot{p}^\mu + \dot{P}^\mu = 0$ or, equivalently, $f^\mu = -\dot{P}^\mu$. Then, since the rest mass of the photon always vanishes, $P_\mu P^\mu = 0$, taking its derivative $d(P_\mu P^\mu)/d\tau = 0$ yields

$$f_\mu P^\mu = 0 \quad (4)$$

which will allow us to build consistent models for the radiation-reaction four-force $(f^\mu) = (f^T, c\vec{f})$ (with $f^T$ the power associated to the thrust $\vec{f}$). All quantities with an arrow on top of them denote spatial three-vectors in Euclidean space. Without loss of generality, one can write down the ansatz $(P^\mu) = E_b/c(\pm 1, \vec{n}_b)$ where $E_b$ stands for the beam energy; the sign $\pm 1$ refers to absorption and emission processes, respectively; and $\vec{n}_b$ is a unit spatial vector pointing in the direction of the beam propagation. Then, one finds from Eq. (4) that

$$f^T = \pm c\vec{f} \bullet \vec{n}_b \quad (5)$$

where $\bullet$ stands for the scalar product with the Euclidean metric between three-vectors. An immediate application of this property arises when one considers the straight motion of a photon rocket, for which $(\lambda^\mu) \equiv \gamma(1, \beta, 0, 0)$ [$\beta = \tanh(\psi)$ with $\psi$ the rapidity] for a motion in the $X$ direction. Then, from Eq. (5), we have that $f^T(\tau) = \pm c f^X(\tau)$ and the equations of motion Eq. (3) now reduce to

$$\begin{cases} \dot{m}c^2 = f^T(\tau)\exp(\mp\psi) \\ mc^2\dot{\psi} = \pm f^T(\tau)\exp(\mp\psi) \end{cases}$$

which can be directly integrated to give

$$m(\tau) = m_0 \exp(\pm\psi). \quad (6)$$

The rest mass is therefore higher (lower) than $m_0 = m(\tau = 0)$ for an absorption (emission) photon rocket (see the above-mentioned sign convention). This result can be put in the following familiar form:

$$\Delta V = c \left| \frac{m^2 - m_0^2}{m^2 + m_0^2} \right| \quad (7)$$

where $\Delta V = \beta c$ is the velocity increase from rest. Equation (7) is nothing but a *relativistic generalization of the Tsiolkovsky equation* for photon rockets, as obtained for the first time by Ackeret in [32] with a different approach. It is important to keep in mind the equivalence of relation (6) between the inertial mass $m$ of the photon rocket and the rapidity $\psi$ and the relativistic Tsiolkovsky equation (7).

With these general elements on photon rockets, we can now build several models of light sails, either perfectly reflecting (white) sails, perfectly absorbing (black) sails, or partially reflecting (gray) sails.

### A. Nonrectilinear motion of a perfectly reflecting light sail

In this section, we will generalize the light sail model of Marx–Redding–Simmons-McInnes [19–21] for rectilinear motion to any arbitrary motion of the light sail. To compute the sail's trajectory from the equations of motion Eqs. (1)–(3), one needs a model for the driving four-force $f^\mu$ that the radiation beam applies on the sail. The propulsion of a reflecting light sail is twofold: first, photon absorption communicates momentum to the sail; second, photon emission achieves recoil of the sail. The case of a perfectly reflecting white light sail corresponds to no variation of the rest mass (internal energy). According to Eq. (3), this happens under the condition that $\lambda_\mu f^\mu_{\text{tot}} = 0$ where the total four-force $f^\mu_{\text{tot}}$ is the sum of the four-forces due to incident and reflected beams $f^\mu_{\text{in}}$ and $f^\mu_{\text{ref}}$. Since we have that $(\lambda^\mu) = \gamma(1, \beta\vec{n}_s)$ (with $\vec{n}_s$ a unit spatial vector pointing in the direction of the sail's motion), the condition of keeping constant the sail's rest mass yields

$$f^T_{\text{tot}} = \beta c (\vec{f}_{\text{in}} + \vec{f}_{\text{ref}}) \bullet \vec{n}_s, \qquad (8)$$

which is different than Eq. (5) since $\vec{n}_s \neq \vec{n}_b$. The perfectly reflecting white sail is indeed a combination of both emission and absorption photon rockets and, thus, Eq. (6) does not hold (the white sail mass is constant) and neither does the relativistic Tsiolkovsky equation (7).

Let us now focus on the time component of the four-force $f^T = dE_{\text{sail}}/d\tau$ (with $\tau$ the sail's proper time) which represents the time variation of the sail's energy $E_{\text{sail}}$ due to the twofold interaction with the beam. The infinitesimal variation of the sail's energy due to the incident beam is given by $dE_{\text{in}} = (IS/c)dw$ where $w = cT - ||\vec{R}||$ is the retarded time, $I$ is the intensity of the beam (in W/m$^2$), and $S$ is the sail's reflecting surface. Indeed, the energy $dE$ that is emitted by the source during the proper time interval $dT$ will reach the sail moving at $\beta = ||\vec{V}||/c = d(||\vec{R}||)/(cdT)$ with some delay ($\beta > 0$) or advance ($\beta < 0$), so that $dE_{\text{in}} = (1 - \beta)dE$ (see also [10] for a spacetime diagram). Therefore, one can write, recalling that $dT = \gamma d\tau$,

$$f^T_{\text{in}} = \gamma \frac{dE_{\text{in}}}{dT} = P(1 - \beta)\gamma \qquad (9)$$

where we set $P = IS = dE/dT$ as the emitted power as measured in the source's frame. Similarly, the infinitesimal variation of the sail's energy due to the reflected beam is given by $dE_{\text{ref}} = (P'/c)du = P'(1 + \beta)dT$ where $u = cT + ||\vec{R}||$ is the advanced time. In this case, the amount of energy $d\mathcal{E}$ that is reflected by the sail during $dT$ is measured by any observer in the source's frame as $dE_{\text{ref}} = d\mathcal{E}(1 + \beta)$, where the factor $(1 + \beta)$ accounts for the sail's motion with respect to this observer. One can thus write

$$f^T_{\text{ref}} = \gamma \frac{dE_{\text{ref}}}{dT} = P'(1 + \beta)\gamma, \qquad (10)$$

where $P' = I'S$ where $P' = d\mathcal{E}/dT$ and $I'$ are the power and intensity of the reflected light beam as measured in the source's frame.

We can now turn on the thrusts $\vec{f}_{(\text{in,ref})}$. Since the total four-momentum of the sail and the beam is conserved during each process of radiation emission and absorption that constitutes the reflection, the thrusts $\vec{f}_{(\text{in,ref})}$ must verify Eqs. (4) and (5), such that we have

$$c\vec{f}_{(\text{in,ref})} = f^T_{(\text{in,ref})} \vec{n}_{(\text{in,ref})}. \qquad (11)$$

From the Snell-Descartes law of reflection, we have that

$$\vec{n}_{\text{in}} \bullet \vec{n}_s = -\vec{n}_{\text{ref}} \bullet \vec{n}_s \equiv \cos(\theta).$$

Putting Eqs. (9)–(11) and the above equation into the constraint of constant rest mass Eq. (8), we find

$$P(1 - \beta)[1 - \beta \cos(\theta)] = -P'(1 + \beta)[1 + \beta \cos(\theta)] \quad (12)$$

together with the following expressions for the components of the four-force:

$$f^T_{\text{tot}} = 2P\gamma\beta \cos(\theta)\left(\frac{1 - \beta}{1 + \beta\cos(\theta)}\right),$$

$$c\vec{f}_{\text{tot}} = P\gamma(1 - \beta)\left(\vec{n}_{\text{in}} - \frac{1 - \beta\cos(\theta)}{1 + \beta\cos(\theta)}\vec{n}_{\text{ref}}\right). \qquad (13)$$

In the particular case of straight motion, one has that $\theta = 0$ ($\vec{n}_{\text{in}} = -\vec{n}_{\text{ref}} = \vec{n}_s$) and the four-force reduces to

$$mc^2 \frac{d\gamma}{d\tau} \equiv f^T_{\text{tot}} = 2P\gamma\beta\left(\frac{1 - \beta}{1 + \beta}\right),$$

$$mc^2 \frac{d(\gamma\beta)}{d\tau} \equiv cf^X_{\text{tot}} = 2P\gamma\left(\frac{1 - \beta}{1 + \beta}\right), \qquad (14)$$

which are the same equations of motion as in [21]. The last of these equations was integrated numerically in [9–11] and also used as a starting point of [13]. However, when the power of the incident radiation beam $P$ is constant, there is a simple analytical solution. Indeed, reexpressing the system Eqs. (14) in terms of the rapidity $\psi$ [$\gamma = \cosh(\psi)$ and $\beta = \tanh(\psi)$] gives the single equation

$$\dot{\psi} = \frac{2P}{mc^2} \exp(-2\psi)$$

the solution of which for a sail starting at rest at $\tau = 0$ is

$$\psi = \frac{1}{2} \log(1 + 4s) \qquad (15)$$

or, maybe more conveniently, in terms of the velocity in the laser's frame,

$$\beta = \frac{V}{c} = \frac{2s}{1 + 2s} \qquad (16)$$

where the dimensionless time $s$ is given by $s = \tau/\tau_c$ with $\tau_c = mc^2/P$ the characteristic time of the light sail travel. For constant power of the incident radiation beam, the light sail takes an infinite amount of time to reach the speed of light, its terminal velocity.

Let us now focus on the nonrectilinear motion of the light sail which is described by the equations of motion with the four-force Eqs. (3) and (13). We recall the ansatz for the tangent vector: $(\lambda^\mu) = \gamma(1, \beta\vec{n}_s)$ with $\vec{n}_s$ a unit spatial vector in the direction of the sail's motion and introducing the rapidity $\psi$ as $\gamma = \cosh(\psi)$ and $\beta = \tanh(\psi)$. Let us consider the $X$ axis as the line between the light source located at the origin of coordinates and the destination and $\vec{e}_X$ as the unit three-vector along the $X$ direction (see also Fig. 1). Due to the Snell-Descartes reflection law, the three-vectors $\vec{n}_s$, the direction of the sail's motion (also normal to the sail's surface $S$), and $\vec{n}_{\text{in,ref}}$ (the directions of the incident and reflected radiation beams, respectively) are coplanars. We can therefore choose the $Y$ axis (with associated unit three-vector $\vec{e}_Y$) so that this plane of reflection corresponds to the $(X, Y)$ plane without loss of generality. The incident radiation beam is emitted from the source and later hits the sail, so that $\vec{n}_{\text{in}} = \vec{R}/||\vec{R}||$ with $\vec{R}$ the sail's position vector. The direction of motion $\vec{n}_s$ does not necessarily point towards the destination and one might have to use this model to compute course correction. Let us denote by $\theta$ the angle of incidence of the radiation beam and the sail's surface, $\vec{n}_s \bullet \vec{n}_{\text{in}} = \cos(\theta) = -\vec{n}_s \bullet \vec{n}_{\text{ref}}$ (where the last equality is due to the Snell-Descartes reflection law), and denote by $\phi$ the angle between the sail's direction of motion and the destination $\vec{n}_s \bullet \vec{e}_X = \cos(\phi)$. We can therefore work with the following ansatz:

$$\vec{n}_s^T = (\cos(\phi), \sin(\phi), 0),$$
$$\vec{n}_{\text{in}}^T = (\cos(\phi - \theta), \sin(\phi - \theta), 0),$$
$$\vec{n}_{\text{ref}}^T = (-\cos(\phi + \theta), -\sin(\phi + \theta), 0),$$
$$\theta = \phi - \arctan\left(\frac{Y}{X}\right). \quad (17)$$

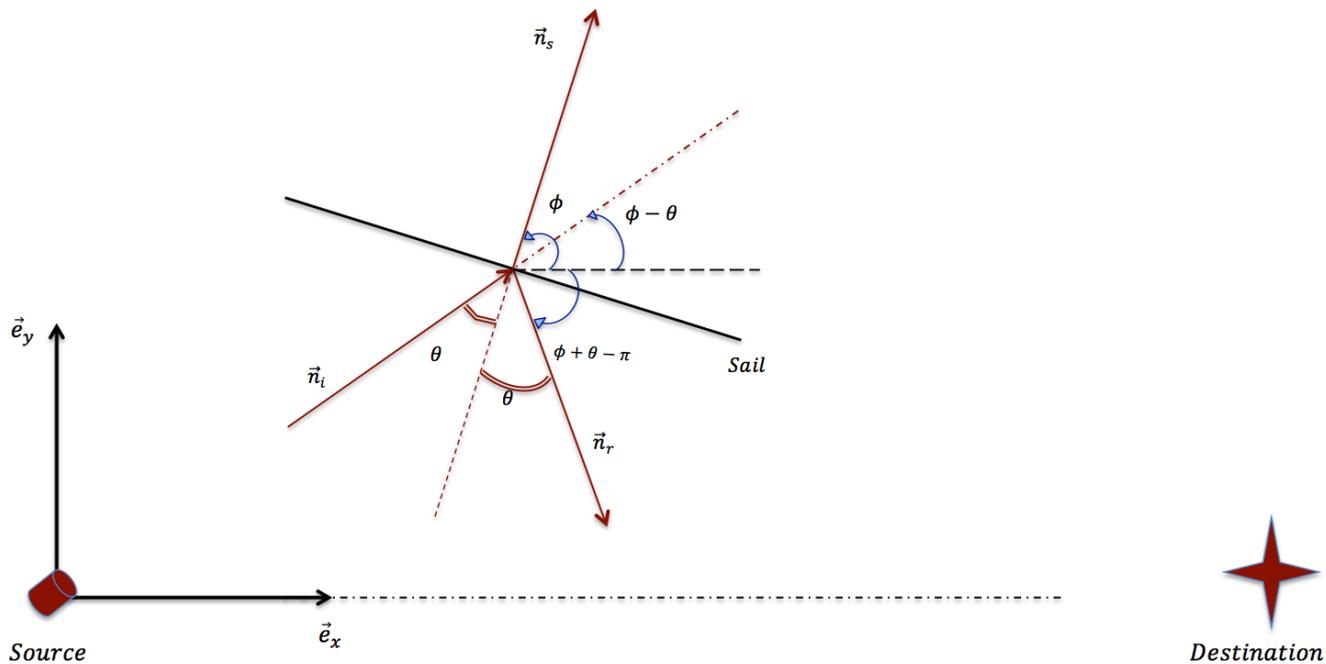

FIG. 1. Sketch of the perfectly reflecting light sail: the light source is located at the origin of coordinates, and the incident beam hits the sail with an angle $\theta$ with respect to the velocity of the sail before it is reflected following the same angle, due to the Snell-Descartes reflection law.

Figure 1 provides an illustration of the triplet of unit vectors $\vec{n}_{s,\text{in,ref}}$ and the angles used.

Thanks to these parametrizations, the equations of motion Eqs. (3) and (13) now reduce to

$$\dot{\psi} = 2\frac{P}{mc^2}\left(\frac{1-\beta}{1+\beta\cos(\theta)}\right)\cos(\theta),$$
$$\dot{\theta} = -\left(\dot{\psi} + \frac{c}{R}\sinh(\psi)\right)\sin(\theta),$$
$$\dot{R} = c\sinh(\psi)\cos(\theta), \quad (18)$$

where $R^2 = X^2 + Y^2$ is the distance from the source to the sail and where $P$, the power of the incident radiation beam, is an arbitrary function. The sail's rest mass $m$ is constant in the case of a perfectly reflecting white sail. These equations will be investigated further in the next section on applications.

### B. Relativistic motion of a perfectly absorbing light sail

The case of a perfectly absorbing black sail corresponds to a perfectly inelastic collision between the photons of the incident radiation beam and the sail, in which the total energy of the beam is converted into both internal and kinetic energy of the sail. As a consequence, the rest mass of the sail is no longer constant. The four-force $(f^\mu) = (f^T, c\vec{f})$ (with $\vec{f}$ the thrust) acting on the sail is now given by the driving power

$$f_{\text{in}}^T = P(1-\beta)\gamma \quad (19)$$

(see previous section) and the thrust

$$c\vec{f} = f_{\text{in}}^T \vec{n}_{\text{in}}, \quad (20)$$

according to Eq. (4). Therefore, the motion of the black sail is rectilinear and directed along the direction $\vec{n}_{\text{in}}$ of the incident radiation beam. Since the black sail is an absorption rocket [Eq. (20) satisfies Eqs. (4) and (5)], its physical properties verify Eq. (6) and the relativistic Tsiolkovsky equation (7).

Without loss of generality, we can identify the $X$ axis to the direction of destination as viewed from the source. The vector $\vec{n}_{\text{in}}^t = \vec{n}_s^t \equiv (\cos(\phi), \sin(\phi), 0)$ is collinear to the sail's position vector and the tangent vector to the sail's world line is given by $(\lambda^\mu) = (\cosh(\psi), \sinh(\psi)\vec{n}_{\text{in}})$ so that the corresponding equations of motion Eqs. (3) can be written:

$$\dot{R} = c\sinh(\psi),$$
$$\dot{m}c^2 = Pe^{-2\psi},$$
$$mc^2 \sinh(\psi)\dot{\psi} = Pe^{-\psi}[1 - \cosh(\psi)e^{-\psi}], \quad (21)$$
$$mc^2 \cosh(\psi)\dot{\psi} = Pe^{-\psi}[1 - \sinh(\psi)e^{-\psi}],$$

with $R$ the Euclidean distance to the source and $\phi = $ const since the motion is rectilinear. Substracting the last two previous equations yields simply $mc^2\dot{\psi} = Pe^{-2\psi}$ and therefore $m(\tau) = m_0 e^\psi$, which is exactly Eq. (6) for an absorption photon rocket, while the rapidity is given by integrating the simple equation

$$\frac{de^{3\psi}}{d\tau} = 3\frac{P}{m_0 c^2} \quad (22)$$

for any given function $P$ modeling the power of the incident radiation beam.

However, one should also consider that part of the photons that has been absorbed by the black sail is thermally reemitted through blackbody radiation, as is done in [10]. Due to the sail's motion, the reemission is anisotropic and produces a Poynting-Robertson drag on the black sail as a reaction force. If one considers that thermal cooling is isotropic in the sail's rest frame, then the four-force for this drag is given by [33]

$$f_{\text{PR}}^T = -P_{\text{rad}}\gamma,$$
$$c\vec{f}_{\text{PR}} = -P_{\text{rad}}\gamma\beta\vec{n}_s \quad (23)$$

where

$$P_{\text{rad}} = \alpha P \frac{1-\beta}{1+\beta} \quad (24)$$

is the power thermally radiated away by the black sail under the form of incoherent radiation. The parameter $\alpha$ is given by $\alpha = 1 - A$ with $A$ the sail's absorptance, so that $\alpha$ describes the part of the incoming radiation power that is thermally reemitted. In [33], dust particles totally reemit the incoming radiation so that it keeps its rest mass $m$ constant, which corresponds to our case $\alpha = 1$. The case $\alpha = 1$ therefore corresponds to blackbody radiation and thermal equilibrium, since then the rest mass stays constant (see below) and so does internal energy and temperature. The case $\alpha = 0$ corresponds to the previous pure black sail without thermal cooling. However, laser-pushed light sails will be hit by tremendous incoming power that might be totally radiated away by thermal cooling, and this could be modeled by $\alpha \neq 1$. This imperfect cooling will make the sails heat and when their thermal capacity is exceeded, they will simply melt. This is therefore an important parameter to take into account.

Accounting both for the radiation pressure from the incoming radiation beam and the Poynting-Robertson drag, the total four-force on the black sail is given by $f_{\text{tot}}^\mu = f_{\text{in}}^\mu + f_{\text{PR}}^\mu$ using Eqs. (19), (20), and (23) and leads to the following equations of rectilinear motion (since $\vec{n}_{\text{in}} = \vec{n}_s$):

$$\dot{m}c^2 = (1 - \alpha)Pe^{-2\psi},$$
$$mc^2\dot{\psi} = Pe^{-2\psi}.$$

These equations can be readily integrated to give

$$m = m_0 e^{(1-\alpha)\psi}, \quad (25)$$

$$\dot{\psi} = \frac{P}{m_0 c^2} e^{-(3-\alpha)\psi} \quad (26)$$

for arbitrary power function $P$.

In the case of constant power $P$, Eqs. (26) can be directly integrated to give the following solutions for the evolutions of the rest mass $m$ and the rapidity $\psi$:

$$m = m_0[1 + (3-\alpha)s]^{\frac{1-\alpha}{3-\alpha}}, \quad (27)$$

$$\psi = \frac{1}{3-\alpha}\log[1 + (3-\alpha)s] \quad (28)$$

with $s = \tau/\tau_c$ the proper time in units of the characteristic time $\tau_c = m_0 c^2/P$ and where we assumed the sail starts at rest at $\tau = 0$. For $\alpha = 0$ (pure black sail without thermal cooling), the velocity in the laser's frame is then simply given by

$$\beta = \frac{V}{c} = \frac{\dot{R}}{c\gamma} = \frac{(1+3s)^{2/3} - 1}{(1+3s)^{2/3} + 1}, \quad (29)$$

while for $\alpha = 1$ (perfect thermal reemission) one finds

$$\beta = \frac{s}{1+s}, \quad (30)$$

which, together with Eq. (16) for the white sail, constitutes a useful set of relations to keep within easy reach for performing estimations. From these results, one can also directly see the differences between a white sail and a black one with thermal cooling. During rectilinear motion, the white sail perfectly reflects backward the incoming radiation, therefore benefiting from twice the incoming momentum to accelerate, while the black sail with Poynting-Robertson drag only benefits from one times the incoming power since it thermally radiates isotropically (in its rest frame). The isotropic thermal reemission of radiation does not give rise to any net acceleration as is the case for white sails in which the reemission is directional.

It must be noticed that, even in the presence of thermal reemission modeled by the Poynting-Robertson drag, the rest mass of the sail is in general not constant. The only exception is for perfect thermal reemission ($\alpha = 1$) for which the mass of the absorbing sail stays constant but then the evolution of the velocity is still different than in the white sail case. This model for the black sail with radiation cooling will be used to model the acceleration phase of the Starshot mission.

### C. General model for arbitrary relativistic motion of a nonperfect light sail

So far we have been considering two extreme cases of a light sail: the perfectly reflecting case (the white sail), which has reflectivity $\epsilon = 1$, and the perfectly absorbing one, which has $\epsilon = 0$. In the first, the rest mass of the sail is constant, while, in the second, inelastic collisions can lead to a variation of the internal energy (rest mass), unless radiative cooling is efficient enough to balance energy absorption. We now need a model for any intermediate values of the reflectivity $\epsilon \in [0, 1]$. Let us therefore write the four-force acting on a gray sail with reflectivity $\epsilon$ as follows:

$$f_g^\mu = \epsilon f_w^\mu + (1 - \epsilon) f_b^\mu \quad (31)$$

where $f_{w,b}^\mu$ stand for the four-forces of the particular cases of a perfectly reflecting (white) sail $\epsilon = 1$ and a perfectly absorbing (black) sail $\epsilon = 0$, respectively. The incoming energy is therefore partly reflected [the white part of the decomposition Eq. (31)], partly converted into rest mass energy (if absorptance $A = 1 - \alpha \neq 0$), and partly thermally radiated away (if $A = 1 - \alpha \neq 1$). Both last effects were included in the black sail model of the previous section. We assume here there is no power transmitted through the sail.

For a perfectly reflecting (white) sail, we have $f_w^\mu = f_{\rm in}^\mu + f_{\rm ref}^\mu$, the composition of incident and reflected beams, with $f_{\rm in}^T = P(1 - \beta)\gamma$, $f_{\rm ref}^T = -P\gamma(1 - \beta)[1 - \beta\cos(\theta)]/[1 + \beta\cos(\theta)]$, and $c\vec{f}_{\rm in,ref} = f_{\rm in,ref}^T \vec{n}_{\rm in,ref}$. For a perfectly absorbing (black) sail with Poynting-Robertson drag, we use $f_b^\mu = f_{\rm in}^\mu + f_{\rm PR}^\mu$ with $c\vec{f}_{\rm in} = f_{\rm in}^T \vec{n}_s$ and $f_{\rm PR}^\mu = -P'_{\rm rad}\gamma(1, \beta\vec{n}_s)$ with $P'_{\rm rad} = \alpha P(1 - \beta)/(1 + \beta)$.

After some simplifications, the equations of motion for the gray sail can be obtained as

$$\dot{m}c^2 = P(1 - \alpha)(1 - \epsilon)e^{-2\psi},$$
$$\dot{\psi} = \frac{P}{mc^2}\left(\frac{1 - \beta}{1 + \beta}\right)\left(1 - \epsilon\frac{1 - \beta\cos(\theta) - 2\cos(\theta)}{1 + \beta\cos(\theta)}\right),$$
$$\dot{\phi} = -2\epsilon\frac{P}{mc^2}\left(\frac{1 - \beta}{1 + \beta\cos(\theta)}\right)\sin(\theta)\cos(\theta). \quad (32)$$

Since the trajectory of the sail is given by $\mathcal{L} = (cT(\tau), X(\tau), Y(\tau), Z = 0)$, finding the trajectory's unknowns $T(\tau)$, $X(\tau)$, and $Y(\tau)$ requires integrating the tangent vector field $\lambda^\mu = dX^\mu/(cd\tau) = (\cosh(\psi), \sinh(\psi)\vec{n}_s)$, with $\vec{n}_s^t = (\cos(\phi), \sin(\phi), 0)$ and $\phi = \theta + \arctan(Y/X)$. It can be checked with some algebra that these most general equations of motion reduce to the previous cases for $\epsilon = 1$ (white sail) or $\epsilon = 0$ (black sail).

We can directly derive an analytical solution for the straight motion of a gray sail. The analytical model below can therefore be used for quickly computing estimations of the rectilinear trajectory. The tangent vector for rectilinear motion along $X$ is given by $(\lambda^\mu) = \gamma(1, \beta, 0, 0)$, such that $\vec{n}_s^T = (1, 0, 0) = \vec{n}_{\rm in}^T = -\vec{n}_{\rm ref}^T$. The four-force acting on the gray sail is given by the decomposition Eq. (31). The white sail component of the four-force is given by

$$f_w^T = 2P\gamma\beta\frac{1 - \beta}{1 + \beta},$$
$$f_w^X = 2P\gamma\frac{1 - \beta}{1 + \beta}$$

from Eq. (14). The black sail component of the four-force is given by $f_b^\mu = f_{\rm in}^\mu + f_{\rm PR}^\mu$, with the Poynting-Robertson drag given by Eqs. (23) and (24). Regrouping all these elements into Eq. (31), one can express the four-force acting on the gray sail in terms of the rapidity $\psi$ as

$$f^T = P\gamma\frac{1 - \beta}{1 + \beta}[1 + \beta + \alpha(\epsilon - 1) - \epsilon(1 - \beta)], \quad (33)$$
$$f^X = P\gamma\frac{1 - \beta}{1 + \beta}[1 + \beta + \epsilon(1 - \beta) - \alpha\beta(1 - \epsilon)]. \quad (34)$$

The equations of motion in this case simply reduce to

$$\dot{m}c^2 = P(1 - \epsilon)(1 - \alpha)e^{-2\psi},$$
$$m\dot{\psi}c^2 = P(1 + \epsilon)e^{-2\psi},$$

which, in the case of constant $\epsilon$ and $\alpha$, give

$$m = m_0 \exp\left(\frac{(1 - \epsilon)(1 - \alpha)}{1 + \epsilon}\psi\right), \quad (35)$$
$$\dot{\psi} = \frac{P}{m_0c^2}(1 + \epsilon)\exp\left(-\frac{3 + \epsilon + \alpha(\epsilon - 1)}{1 + \epsilon}\psi\right) \quad (36)$$

for arbitrary variable power $P$. In the simplest case of constant power $P$, the following analytical solution can be found:

$$m = m_0\{1 + [3 + \epsilon + \alpha(\epsilon - 1)]s\}^{\frac{(1-\epsilon)(1-\alpha)}{3+\epsilon+\alpha(\epsilon-1)}},$$
$$\psi = \left(\frac{1 + \epsilon}{3 + \epsilon + \alpha(\epsilon - 1)}\right)$$
$$\times \log\{1 + [3 + \epsilon + \alpha(\epsilon - 1)]s\} \quad (37)$$

for a gray sail with reflectivity $\epsilon$ and absorptance $1 - \alpha$, starting from rest with rest mass $m_0$ ($s = \tau/\tau_c$ with $\tau_c = m_0c^2/P$). These results are consistent with the previously obtained particular solutions for $(\epsilon, \alpha) = 1$ [Eq. (15)] and $(\epsilon, \alpha) = 0$ [Eqs. (27) and (28)].

## III. APPLICATIONS

We propose here three applications of the original light sail model derived in the previous section. First, we present the general dynamics of perfectly reflecting light sails, then we apply our model to fly-by missions at relativistic velocities, and finally we apply our model to a single trip with double-stage light sails.

### A. Sailing at relativistic velocities

Let us consider the general motion of perfectly reflecting white light sails, as given by Eqs. (18). In this model, the sail is reflective on both sides of its surface $S$. Figure 2 presents several trajectories of light sails coming from infinity at $\tau \to -\infty$ and passing closest to the laser source at $\tau = 0$ with distance $R_0 = R_{\min}$ for different velocities $\beta_0$ and inclination $\theta_0$. After the closest approach, the sail is deflected by the light source depending on the sail's velocity at closest approach $\psi_0$ and the sail's inclination $\theta_0$ there. A complete study of the light sail's phase diagram is given in Figs. 3 and 4. First, this system admits only one fixed point ($\theta = \pi/2, \psi' = 0$), which is an unstable saddle point. Indeed, a linear stability analysis of system (18) indicates that the Jacobian of the right-hand side has two eigenvalues of opposite signs ($\lambda_{1,2} = (1 \pm \sqrt{1 + 4/R})/2$; $R \geq 0$). This unstable equilibrium point is surrounded by two attractors ($\theta \to$ $0, \psi' \to 0)$ and $(\theta \to \pi, \psi' \to -2)$ [$(\theta \to \pi, \psi' \to 0)$ for power decaying as $1/R^2$]. Due to this configuration of the phase space, there are trajectories starting and ending with $\vec{n}_s$ collinear ($\theta = 0$) or anticollinear ($\theta = \pi$) to the incident beam for vanishing velocities $\beta_0$ at closest approach (loop-shaped trajectories drawn with straight lines in Figs. 3 and 4). But if velocity at closest approach $\beta_0$ is large enough trajectories can overcome the unstable fixed point and transit from the vicinity of one attractor to the other. This is the case of trajectories shown in dashed lines in Figs. 3 and 4.

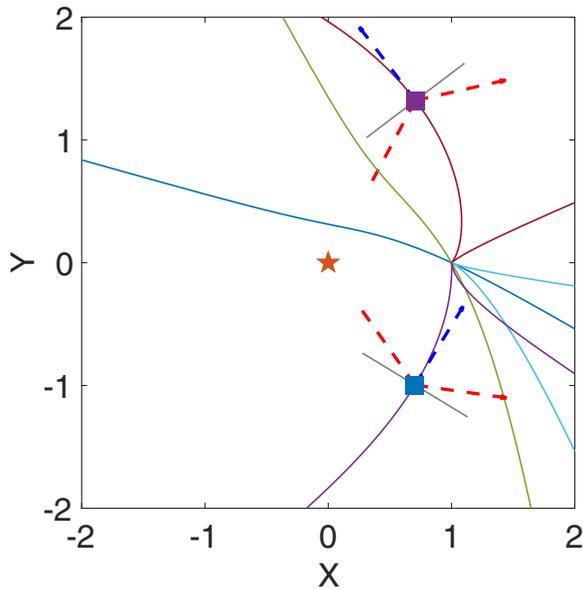

FIG. 2. Various trajectories of perfectly reflecting white sails with driving power $P$ decaying as $1/R^2$. A star indicates the location of the laser source, and a square indicates one particular position of the light sail with an associated reflecting surface and vector triplets $\vec{n}_{s,\text{in,ref}}$.

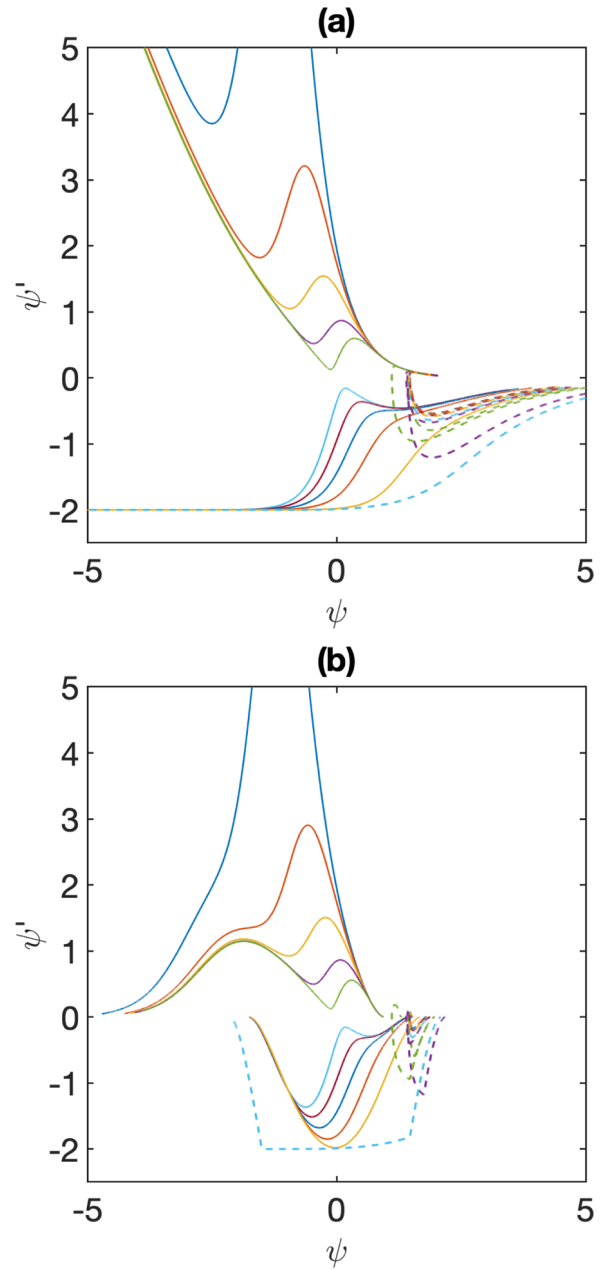

FIG. 3. Phase diagram in the plane $(\psi, \psi')$ of the perfectly reflecting white sail. The upper plot is for constant power $P$ while the lower plot shows the case of driving power $P$ decaying as $1/R^2$. Straight lines correspond to $\beta_0 = 0$ and dashed lines correspond to $\beta_0 = 0.9$.

This singular configuration of the phase space, with two attractors surrounding an unstable equilibrium, makes laser sailing at relativistic speed an intricate and delicate discipline.

### B. Acceleration phase of a nanoprobe interstellar mission

Let us apply our results to the Starshot mission [9–11,13,23]. This project aspires to send tiny light sails of 1-g mass-scale towards Proxima Centauri for a fly-by at a cruise velocity of about 20% that of light. The extreme kinetic energy would be provided by a gigantic ground-based laser during an acceleration phase lasting a few hours. The probes will then freely fly toward Proxima Centauri, which they should reach within about 20 years. Let us therefore consider a rest mass at start $m_0$ of 1 g and a powerful laser source emitting light with power $P_0 = 4$ GW. We also assume, following [11], that the size of the laser source is 10 km and that of the sail is 10 m so that is the maximal distance. If the laser's wavelength is 1064 nm, then the maximum distance $D_{\text{max}}$ up to which the laser beam completely encompasses the sail is about 0.3 A.U. The characteristic time $\tau_c = m_0 c^2/P_0$ of this system is about 6.2 h. We can now study the acceleration phase of such spacecrafts starting from rest as a function of the sail's reflectivity $\epsilon$, with the model derived in the previous section (including Poynting-Robertson drag).

In order to determine the gray sail's trajectory, one needs to integrate the tangent vector field $\lambda^\mu$ and use Eqs. (32). One must also complete this set of equations by a model for the power of the incident radiation beam $P$ as well as for the sail's absorptance $A = 1 - \alpha$, which are free functions in our formalism. For the driving power $P$, we will consider here the

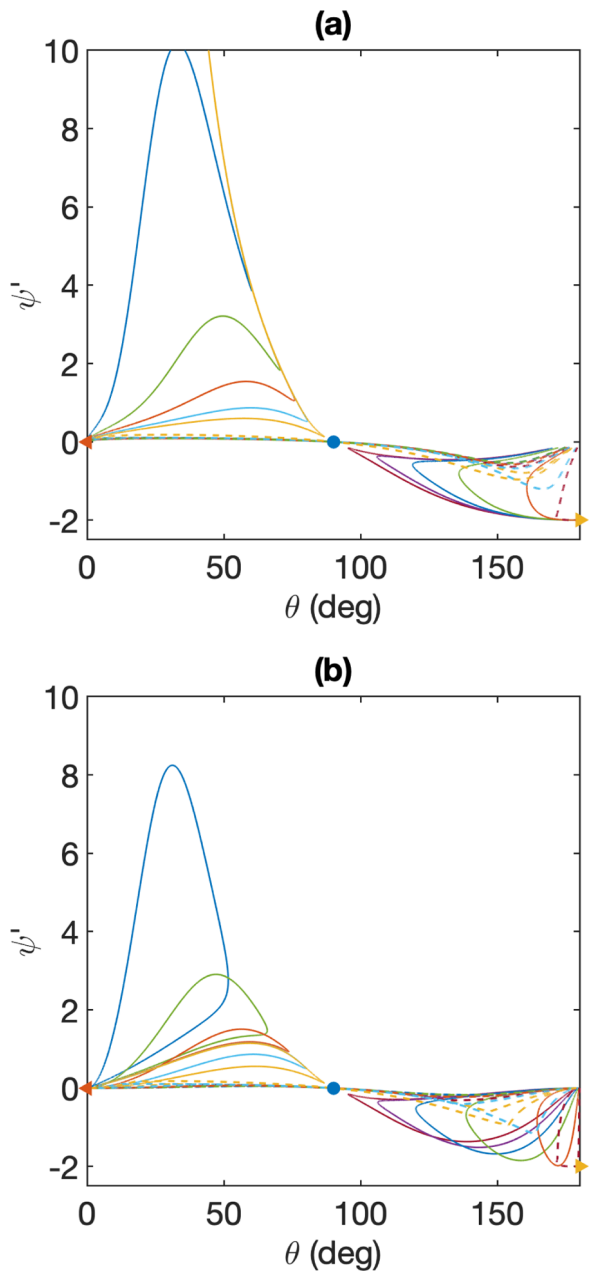

FIG. 4. Phase diagram in the plane $(\theta, \psi')$ of the perfectly reflecting white sail. The upper plot is for constant power $P$ while the lower plot shows the case of driving power $P$ decaying as $1/R^2$. The dots and the triangles, respectively, indicate the location of the unstable saddle point $(\theta = \pi/2, \psi' = 0)$ and the attractors $(\theta \to 0, \psi' \to 0)$ and $(\theta \to \pi, \psi' \to -2)$ [$(\theta \to \pi, \psi' \to 0)$ for decaying power]. Straight lines correspond to $\beta_0 = 0$ and dashed lines correspond to $\beta_0 = 0.9$.

following simple model, which was introduced in [11]

$$P = \begin{cases} P_0; & D < D_{\max} \\ P_0 \left(\frac{D_{\max}}{D}\right)^2; & D \geqslant D_{\max} \end{cases} \quad (38)$$

where $D(\tau) = [X^2(\tau) + Y^2(\tau) + Z^2(\tau)]^{1/2}$ is the time-dependent Euclidean distance between the source of the propelling radiation and the sail and $D_{\max}$ is the maximum distance at which the beam spot encompasses the sail's surface $S$. This maximum distance is related to the sail's characteristic size $R$; the one of the beam source, $r$; and the radiation wavelength $\lambda$ by the following relation (see [11]):

$$D_{\max} \approx \frac{rR}{2\lambda}$$

up to some order of unity geometrical factor depending on the shape of the beam source. In this model, the energy that propels the sail beyond the distance $D_{\max}$ decays as the inverse of its distance to the source. A more sophisticated but also more realistic model for the beam power is the Goubau beam of [10], the shape of which is qualitatively similar to the simple model of [11] used here.

In order to determine a model for the absorptance $A = 1 - \alpha$, one must recall that the radiated power $P_{\mathrm{rad}} = (1 - \epsilon)\alpha P(1 - \beta)/(1 + \beta)$ is related to the sail's temperature $T$ by the Stefan-Boltzmann law: $P_{\mathrm{rad}} = \sigma S T^4$ with $\sigma \approx 5.67 \times 10^{-8}$ W/(m$^2$ K$^4$) and $S$ the sail's surface. The absorptance is therefore varying with time, as driving power $P$ decreases and the sail's velocity increases with the distance. The variation of the sail's temperature is also related to that of rest mass (internal energy) by

$$\dot{T} = \frac{c^2}{C_V}\dot{m}$$

with $C_V$ the sail's specific heat. Deriving Stefan-Boltzmann's law with respect to proper time $\tau$ leads one to write down an additional differential equation for $\alpha$:

$$\dot{\alpha} = \alpha \mathcal{P}\left(2\mathcal{P}\dot{\psi} + \frac{d\mathcal{P}}{dR}\sinh(\psi)\right) \\ + \cdots 4\mathcal{C}\mathcal{P}^{7/4}\left(\frac{1+\beta}{1-\beta}\right)^{1/4}\alpha^{3/4}\dot{m} \quad (39)$$

where $\mathcal{P} = P/P_0$, $R$ is the sail's distance to the source, and

$$\mathcal{C} = \left(\frac{\sigma S}{P_0(1-\epsilon)}\right)^{1/4}\frac{c^2}{C_V}.$$

Equation (39) completes the system Eqs. (32).

The first point we propose to address is the evolution of the absorptance during the very early phases of acceleration. Since the constant $\mathcal{C}$ is large, for a specific heat $C_V$ of order 2 kJ/(kg K) (like that of graphene [34] at 1500 K, see below), a sail's surface of $S = 16$ m$^2$ (see also [10]), and a reflectivity $\epsilon$ of 99.9%, one has that $\mathcal{C} \approx 10^{10}$. The last term in the right-hand side of Eq. (39) is therefore dominant, leading to a very fast evolution of $\alpha$. Starting at rest from a sail's initial temperature of 100 K, one has that $\alpha(\tau = 0) \approx 10^{-5}$ and quickly evolves toward vanishing absorptance $\alpha \to 1$ where the sail's thermodynamic state converges towards thermal equilibrium (constant temperature and rest mass). Figure 5 represents the evolution of both the sail's absorptance $1 - \alpha$ and temperature over a short time range of $10^{-2}$ s with the specifications given above. The value of the absorptance $A = 1 - \alpha$ reached after $10^{-2}$ s is around $10^{-8}$, the temperature is around 1500 K, and the variation of internal energy is around 2.5 MJ. In what follows, we will therefore assume vanishing absorptance $\alpha = 1$ in our simulations over a longer duration.

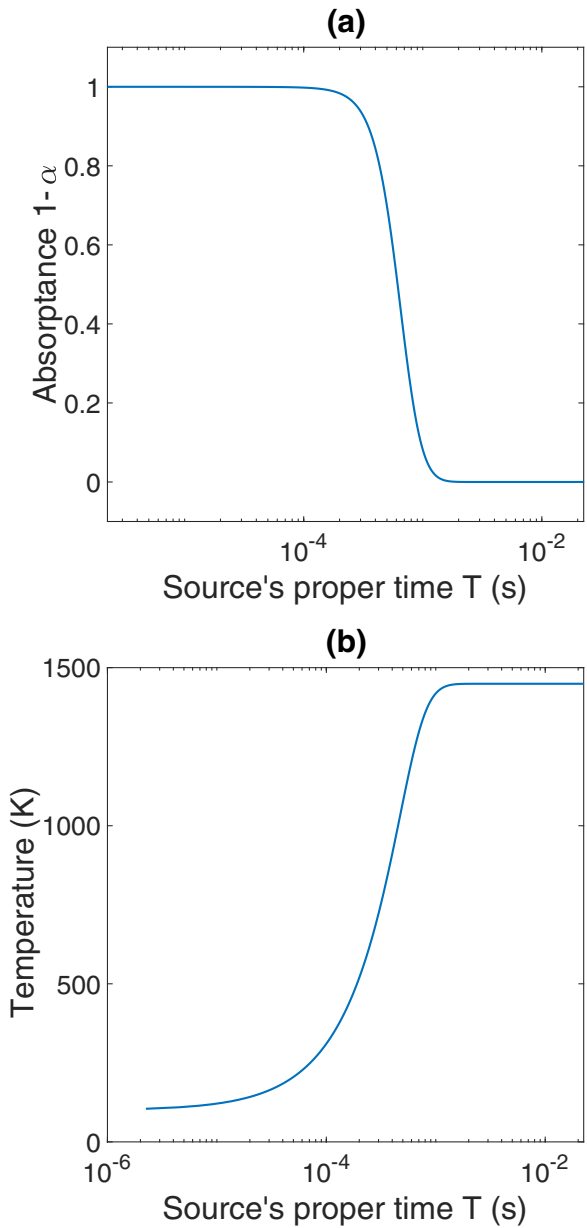

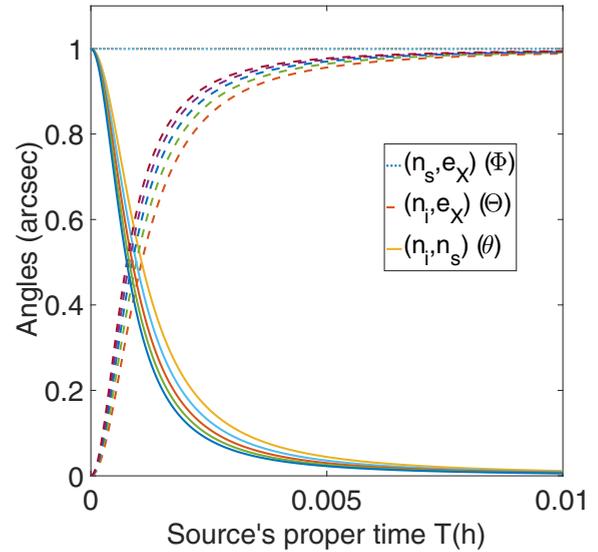

FIG. 6. Evolution of the angles $\theta$ (straight line), $\phi$ (dotted line), and $\Theta$ (dashed line) in the very early phases of acceleration for five values of $\epsilon$ equally spaced between 0 and 1.

FIG. 5. Evolution of the absorptance $1 - \alpha$ (a) and of the sail's temperature (b) for the following parameters: $P0 = 4$ GW, $S = 16$ m$^2$, $C_V = 1900$ J/(kg K), and a reflectivity $\epsilon$ of 99.9%.

We now move on to consider the aiming accuracy to reach such a far-away destination such as Proxima Centauri. Indeed, at the start the light sails might not be perfectly perpendicular to the incident radiation beam or to the direction of destination. As before, let us denote by $\theta$ and $\phi$ the angles between the vector $\vec{n}_s$ normal to the sail's surface and the incident radiation beam $\vec{n}_{in} = \vec{R}/||\vec{R}||$ and the direction $e_X$ from the laser's source to the destination. One last angle is $\Theta$, which gives the position of the light sail with respect to the destination. One piece of good news is that any small misalignment will be quickly corrected naturally during the acceleration phase. Indeed, Fig. 6 gives the evolution of the three angles $\theta$, $\phi$, and $\Theta$ characterizing the gray sail's dynamics. One can see that the initial misalignment $\theta_0 = 1''$ (1 arc sec) quickly vanishes after the start of the acceleration phase. This is due to the vicinity of the attractor ($\theta = 0$, $\psi' = 0$) presented in the previous section. For perfectly reflecting light sails ($\epsilon = 1$), the Snell-Descartes reflection law imposes $\phi = \Theta + \theta$ at all times. Since $\theta$ goes to zero and $\phi$ remains almost constant, this yields $\Theta \to \phi \approx \theta_0$ so that the sail's velocity (angle $\phi$) quickly aligns with the incident radiation beam (angle $\Theta$) and the sail's trajectory becomes radial, in the direction $\Theta \approx \theta_0$. As a consequence, the transverse deviation at the destination is given by $Y \approx R_{des}\theta_0$ with $R_{des}$ the distance between the source and the destination. Therefore, an initial misalignment $\theta_0$ of only 1 arc sec simply results in the case of Starshot in a deviation of 81 A.U. after a $R_{des} = 4.4$-light-year trip. Our model allows also accounting for an initial error in aiming to the destination, i.e., $\Theta_0 \neq 0$, which leads to the same asymptotic behavior: $\theta \to 0$ and $\Theta \to \phi$.

Since any small initial misalignment $\theta_0 \ll \pi/2$ is swept away by the driving force, the sail's trajectory shortly becomes close to rectilinear. Figure 7 represents the evolution of velocity in the source's frame, proper acceleration $a \approx cd\beta/d\tau$ felt by the light sail, and distance $R$ to the source for various values of its reflectivity $\epsilon$. The velocity quickly saturates once the sail overcomes $D_{max}$ and the driving power decays with the inverse square of distance. It must be pointed out that the Starshot space probes will experience an effective gravitational field (given by proper acceleration) as large as 2500 times that of Earth during the first hour of acceleration toward their cruise velocity. During this extreme first hour they will cross the distance to $D_{max}$, of about 1/3 A.U., and they will reach a distance equivalent to that of Jupiter in only 4 h. The energy cost spent by the driving source $E_0 = P_0 T$ (with $T \approx 4$ h the total duration of the acceleration phase) to achieve that is about $5.4 \times 10^{13}$ J or 12.9 kt (the equivalent of one Little Boy class A bomb) per Starshot probe. One can therefore ask what is the efficiency of such an energetic waste. This is given by examining the ratio of the total kinetic energy $E_K = (m\gamma - m_0)c^2$ communicated to each probe to the source's energy cost $E_0$. The efficiency is about 3.4% for white sails ($\epsilon = 1$) and 1.8% for black sails at thermodynamic equilibrium ($\epsilon = 0$, $\alpha = 1$).

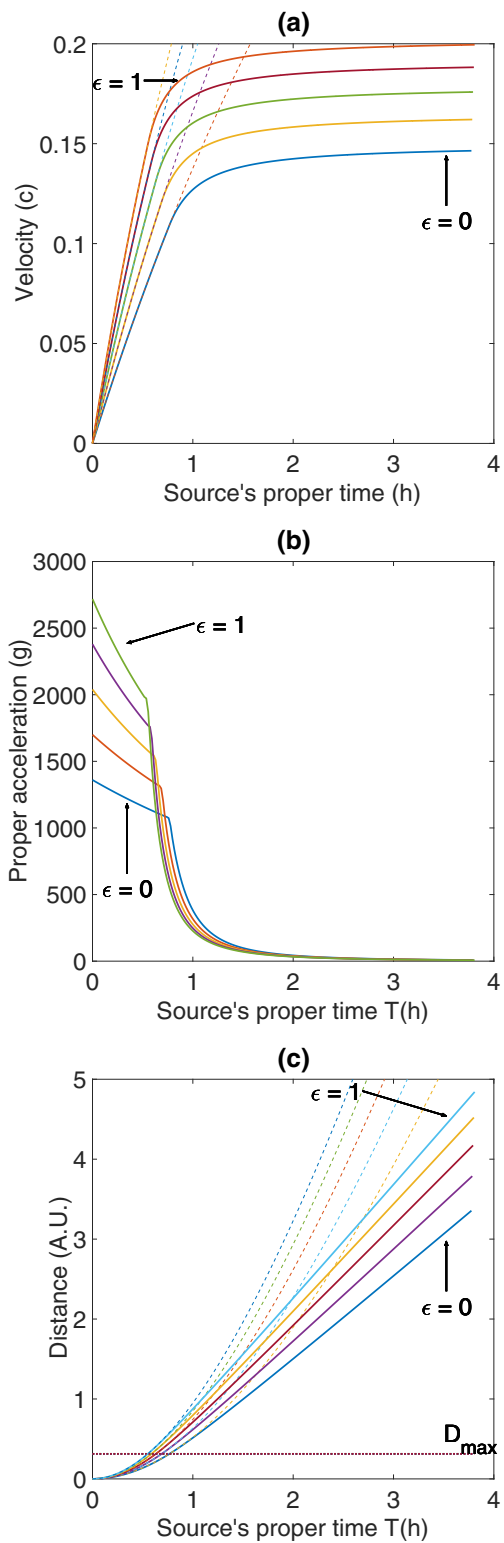

FIG. 7. Evolution of the light sail's velocity (a), proper acceleration (b), and distance to the source (c) with time for five values of $\epsilon$ equally spaced between 0 and 1. Straight lines give the results for a decaying driving power given by Eq. (38) while dashed lines are for constant power.

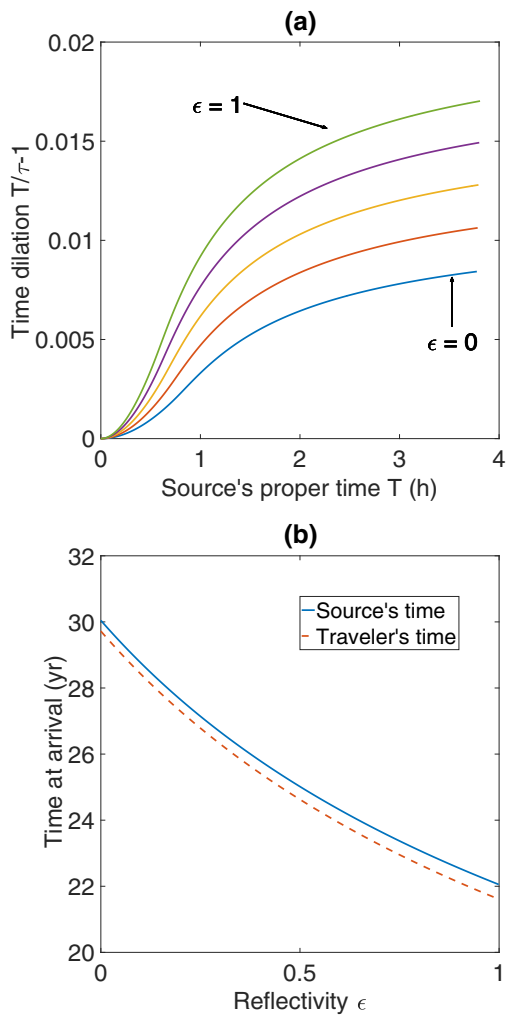

FIG. 8. (a) Evolution of time dilation $T/\tau - 1$ aboard during the acceleration phase for various sails' reflectivity $\epsilon$. Straight lines indicate the prediction for constant power. (b) Trip duration of the Starshot probes towards Proxima Centauri as a function of the sail's reflectivity $\epsilon$.

Almost perfectly reflecting white sails ($\epsilon \approx 1$) have also the advantages of shorter trip duration toward Proxima Centauri and lower temperature at thermodynamic equilibrium, should they survive the acceleration phase and the trip (see also [13]).

As shown in Fig. 8, white sails ($\epsilon = 1$) arrive at a destination nearby Proxima Centauri in about 22 years against 30 years for black sails ($\epsilon = 0$), from the point of view of mission control on Earth. Indeed, time aboard these relativistic probes will also elapse slower, by about 2% for white sails as also shown in Fig. 8. Once integrated over the whole duration of the mission, this represents almost a difference of 162 days between on-board time and mission control time at the end of the mission (22 years for $\epsilon = 1$). From the point of view of the Starshot probes, this relativistic effect will shorten the effective trip duration by 162 days. This effect must be taken into account to wake up on-board instrumentation and start the scientific program at the right local time, otherwise the destination system will be missed by almost $5.6 \times 10^3$ A.U. ($= 0.2c \times 162$ days).

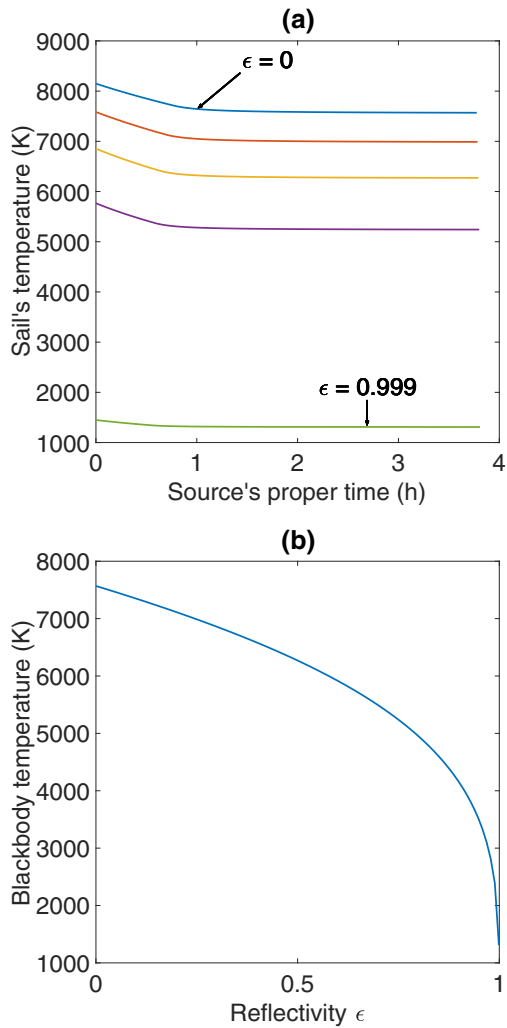

FIG. 9. Evolution of the Starshot probe's blackbody temperature during acceleration phase (a) and final temperature at the end of the acceleration phase as a function of the sail's reflexivity $\epsilon$ (b).

We can now conclude this analysis of the Starshot project by giving the evolution of the Starshot probe's temperature at thermodynamic equilibrium, as is done in Fig. 9. Blackbody temperature quickly stabilizes during the acceleration phase as a result of driving power decay and convergence towards terminal velocity. The final temperature at the end of acceleration depends nonlinearly on the sail's reflectivity $\epsilon$: while around 1500 K above 99.9% reflectivity, it reaches 5500 K at 70% and almost 8000 K at 0%. These results strongly motivate the investigation of carbon-based structures like graphene [for which the specific heat at 1500 K is around 1.8 kJ/(kg K)] [34] or metamaterials allowing one to manipulate thermal radiation [35] for the design of Starshot's extreme sails.

### C. Single trips in the Solar System with two-stage light sails

In the context of large scale directed energy propulsion, several interesting proposals for the exploration of the Solar System at sublight velocities have been made (see, for instance, [14]). Among these, the idea of reconverting on-board the laser's illumination to drive an ion thruster leads to an efficient solution for high-mass missions. We do not pretend to propose a detailed Solar System mission here but rather we revisit Forward's idea of multistage light sails with the tools established in this paper.

The major issue with laser-pushed light sails, according to the pioneer Marx in [19], was the slow-down once approaching the destination since this external propulsion cannot be reversed. That is why Forward suggested in [5] to use multistage light sails: the large sail that is used during the acceleration phase separates at some point between an outer ring and a central breaking sail to which a payload is attached. After separation, the laser source is still propelling the outer sail, that stays on an accelerated trajectory while the payload reverses its own sail to catch the reflected light coming from the outer ring and uses its radiation pressure to reduce its speed.

We propose to revisit this idea with our model by giving a detailed trajectory of a single trip inside the inner Solar System, beyond the simple description made in [5]. For the sake of simplicity, we will restrict ourselves to rectilinear motion. The abscissas of the outer ring and payload $X_{o,p}$ indicate their distance to the laser source. To determine the trajectory of the double-stage light sail, we make use of equations of motion Eqs. (35) and (36). During the acceleration phase, the inner payload sail and the outer ring sail are bound and evolve together at distance $X_o = X_p$ and with total rest mass $m_{tot} = m_o + m_p$. The light source illuminates the outer sail with power $P_o$ given by the following function [see also Eq. (38)]:

$$P_o = \begin{cases} P_0; & X_{o,p} < X_1 \\ P_0\left(\frac{X_1}{X_{o,p}}\right)^2; & X \geqslant X_1 \end{cases} \quad (40)$$

where $P_0$ is the power emitted by the light source located at the origin of coordinates. $X_1$ stands for the maximal distance at which the light source completely encompasses the outer sail and after which the illumination decreases as the inverse of the distance squared. At some point the outer ring sail separates from the inner payload ring, which now uses its front reflective surface to collect light reflected from the outer sail to decelerate. The payload sail then enters a deceleration phase which is driven by the following power function:

$$P_p = \begin{cases} -P_o; & (X_o - X_p) < X_2 \\ -P_o\left(\frac{X_2}{X_o - X_p}\right)^2; & (X_o - X_p) \geqslant X_2 \end{cases} \quad (41)$$

where $P_o$ is the power acting on the outer sail Eq. (40), $X_2$ is the maximal distance at which the light reflected from the sail completely covers the payload's sail located at $X_p$, and $X_o \geqslant X_p$ is the position of the accelerating outer sail.

Figure 10 presents a typical example of a rectilinear single trip performed with two-stage light sails. We consider a payload of rest mass $m_p = 20$ tons and an outer sail of mass $m_o = 4$ tons (hence a total mass of 24 tons) powered by a laser source of power $P_0 = 10$ GW. Following [11], if we assume a sail of thickness 1 $\mu$m and density 1.4 g/cm$^3$, then the radius of the outer sail with a mass of 4 tons is $\approx$954 m. We also assume the laser source is capable to illuminate the outer sail with constant power up to a distance $X_1$ of 5 A.U. after which the illumination starts decreasing with the inverse square of the distance to the source. Similarly, the outer sail is able to completely illuminate the inner payload sail up to a relative distance $X_2 = X_o - X_p$ which we choose to be equal to 5 A.U. in the example of Fig. 10. The light source is used to both accelerate and decelerate the double-stage spaceship, which means that it must completely illuminate the outer sail over a large distance (here $X_1 = 5$ A.U.) covering both deceleration and acceleration. If the outer sail has a radius of 954 m as we have seen above, this could be achieved with a directed energy system of radius close to 2 km, assuming an infrared laser of wavelength equal to 1064 nm (Nd:YAG laser, see also [11]).

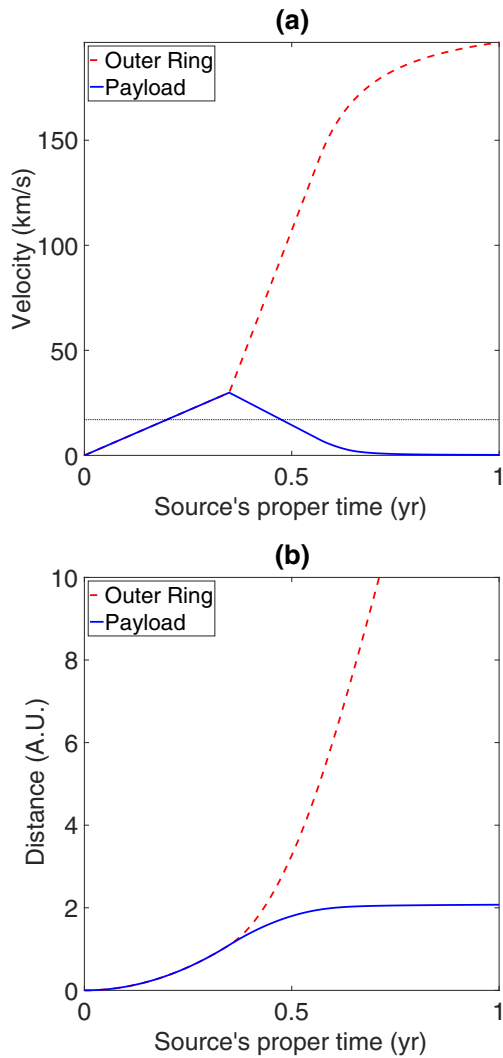

FIG. 10. Evolution of the outer and inner payload sails' velocities (a) and distance (b). A dotted line indicates the cruise velocity of the Voyager 1 probe.

After the separation, the lighter outer sail is freed from its heavier inner payload sail and therefore undergoes a stronger acceleration and a quickly increasing relative distance to the payload. This implies that the duration of the breaking phase is shorter than the acceleration one and that the decay of the illumination of the inner payload sail arrives earlier as the outer sail quickly goes out of the distance $X_2$ at which the light from the outer sail completely illuminates the inner payload sail. The lighter the outer sail the more important its postseparation acceleration and the more important this effect.

In the example illustrated in Fig. 10, the acceleration phase of the double-stage 24-ton spaceship lasts for 4.2 months, after which it reaches a cruise velocity of about 30 km/s, almost doubling the current record of 17 km/s held by the Voyager 1 probe. The separation of the payload from the outer ring sail then occurs and a breaking phase decelerates the payload for about three months. After that, the illumination from the outer sail quickly fades away, and the outer sail has reached a velocity of about 160 km/s and a distance $X_o$ of 7 A.U. which is $X_2 = 5$ A.U. further than the distance reached by the payload. It takes about seven months for the 20-ton payload to reach a distance of 2 A.U., roughly the average distance separating planets Earth and Mars. The total energy spent by the laser source to make this trip possible is huge, around $2 \times 10^{17}$ J or 45 000 kt. The efficiency of the directed energy system at the end of the acceleration phase is very poor, close to $10^{-4}$.

## IV. CONCLUSION

Among a true mess of various ideas for making interstellar travel possible, some were not as fanciful as could have appeared on first sight. In fact, interstellar travel might well have been invented almost 60 years ago by Forward [6] and Marx [19]. Their vision was to use the then freshly realized laser to propel reflecting sails at relativistic velocities, towards the stars. This rather old idea has taken a long time to be fully explored within relativity. Curiously, Forward, although a renowned specialist of this discipline, did not push his idea very far into formalism and detailed computations in the relativistic limit in his papers on the subject. A relativistic model for the straight motion of a perfectly reflecting light sail was introduced by Marx in 1966 [19], then seriously corrected by Redding in 1967 [20] before being revisited into its final form in 1992 by Simmons and McInnes in a pedagogical paper. Under the recent burst of interest accompanying NASA's Starlight, Breakthrough Initiative, and other programs in the previous two decades, this restricted model has served as a basis for a variety of interesting extensions [7,9–11,14,24]. Unfortunately, these attempts did not revise the fundamentals of the Marx–Redding–Simmons-McInnes model into special relativity, which lead to some incompleteness of the recent models and sometimes confusing presentations. For instance, one major drawback of these recent papers is that they did not explore the case of nonpure forces. By doing this, we have been able to provide a model for predicting the sail's temperature at thermodynamic equilibrium. In addition, the nonrectilinear motion of a light sail has not been investigated so far.

This paper introduced the appropriate formalism to go beyond this situation, by deriving the general model for the nonrectilinear motion of partially reflecting gray light sails starting from general principles in special relativity. As part of the family of photon rockets, the light sails have to deal with four-forces that obey several constraints and exhibit several interesting properties, including the variation of their rest mass when inelastic collisions with the incoming radiation occur. The general model of gray sails is built on a combination of two particular cases: first, the one of perfectly reflecting white sails which have a general planar motion; second, the one of perfectly absorbing black sails the motion of which is along the direction of the incoming radiation beam and ruled by the push of external radiation pressure and the drag force from the Poynting-Robertson effect. Our model requires numerical integration although in simplistic cases analytical solutions allow crude approximations.

We also presented three applications of our model. First, sailing at relativistic velocities has been shown to be intricate due to the instability of the equilibrium when the sail stands parallel to the incoming beam. Whatever the initial conditions outside of this unstable point, any light sail will naturally relax toward two possible positions with the velocity anti- or colinear with the direction of the incoming beam both with $c$ as asymptotic speed. This behavior will allow light sails to spontaneously align with the incoming beam while accelerated, yielding a prediction for a deviation from their initial aim. Second, we provide the predictions of our model on the Starshot mission, within some generic nonrestrictive assumptions that could be easily adapted if desired. The model accounts for initial deviations of the sail's position and velocity vectors from the direction of the destination, which produces non-negligible deviations at arrival. To show this, we have provided a simple example with a small initial misalignment of 1-arc-sec amplitude, and our predictions on the light sail's velocity, proper acceleration, and distance are comparable to those of previous works [9–11], although accounting for transverse deviation, rest mass, and temperature variation when the sail is partially absorbing. There are several important points for mission design that have been derived here: aiming accuracy, proper acceleration and mechanical constraints, increase of the sail's temperature, time dilation effect and wake-up time of the probe's internal systems, and propulsion efficiency and the need to seriously consider power recycling to improve it. Third, we provide a very simple model of a single trip in the Solar System with a ton-scale two-stage light sail and a gigawatt-scale laser, to illustrate the potential interest of this technique for interplanetary exploration.

Of course, the results presented here are not exhaustive and should be carefully extended to a more realistic numerical modeling prior to any directed energy mission. While our model is valid for nonrectilinear motion and general sail reflectivity and absorptance, there are several physical effects that must be added to compute a precise trajectory of such an intricate (and costly) mission towards far-away planets or even stars. This includes modeling of the environment: incoming beam properties and interaction with the sail, gravitational influences, Solar System constraints (e.g., a free path from a ground-based laser and a sail), interplanetary and interstellar media, magnetic fields, and anisotropic thermal radiation, to name but a few. Several papers have already started investigating some of these effects (see, for instance, [10,13] for a review), which could now benefit from the general formalism presented here. Other concerns of the authors are telecommunication and astronavigation issues which must take into account several effects from general relativity (see [12] for a first approach in the case of straight motion of photon rockets). Following [10], we agree that a serious effort of physical and numerical modeling, beyond simplistic models sometimes solved using spreadsheets, must be made prior to any launch. We also think that small scale ground-based experiments and flying prototypes will be necessary to calibrate the models and adjust numerical simulations (see [10] for an interesting suggestion).

The key idea of Forward and Marx that could well make interstellar travel possible one day is undoubtedly the externalization of the propulsion's energy source. As Forward already envisions in [5], the most obvious primary energy source for that purpose is the almost inexhaustible amount of energy radiated by the Sun. But this means it has to be collected by huge amounts, probably on a planetary scale, and converted into intense radiation beams that will be efficient for propulsion but potentially also for other dangerous aims. The sending of one single tiny probe, of gram mass scale, toward Alpha Centauri individually requires as much energy as the one delivered by several kilotons of TNT. Besides this frightening energy waste, cost estimation for such a program is of the order of $10 \times 10^9$ dollars [10]. Of course, it is natural to wonder whether the stars are worth these sacrifices, no matter their appealing beauty to the stargazer. In many respects, like exploration of the unknown and maybe one day our own survival, they certainly are. Yet, it is a charming idea that the key to reach the stars could lie in the glare of their close and lovely yellow dwarf cousin.

## ACKNOWLEDGMENTS

The authors thank T. Carletti and M. Lobet for useful discussions as well as both referees for their constructive reports that enabled us to improve the paper. In particular, the enthusiasm of the referees—which was very welcome in these hard times—has encouraged us to pursue our research on this fascinating subject.